\documentclass[%
aps,
reprint,
prl,
superscriptaddress,
nofootinbib,
nobibnotes,
showkeys,
floatfix,
]{revtex4-2}

\usepackage{amsmath,amssymb,bm}
\usepackage{graphicx}
\usepackage[caption=false]{subfig}
\usepackage[percent]{overpic}
\usepackage[%
colorlinks=true,
linkcolor=blue,
citecolor=blue,
]{hyperref}

\begin{document}

\title{Critical Scalarization for a Self-Gravitating Bosonic Condensate}

\author{Meng-Yun Lai}
\email{mengyunlai@jxnu.edu.cn}
\affiliation{School of Physics, Jiangxi Normal University, Nanchang 330022, China}

\author{Hyat Huang}
\email{hyat@mail.bnu.edu.cn}
\affiliation{School of Physics, Jiangxi Normal University, Nanchang 330022, China}

\author{Wen-Cong Gan}
\email{ganwencong@jxnu.edu.cn}
\affiliation{School of Physics, Jiangxi Normal University, Nanchang 330022, China}

\author{De-Cheng Zou}
\email{dczou@jxnu.edu.cn}
\affiliation{School of Physics, Jiangxi Normal University, Nanchang 330022, China}

\author{Yun Soo Myung}
\email{ysmyung@inje.ac.kr}
\affiliation{Center for Quantum Spacetime, Sogang University, Seoul 04107, Republic of Korea}

\begin{abstract}
We establish a minimal nonrelativistic realization of scalarization in a self-gravitating bosonic condensate coupled to a scalar response field.
A local effective-mass shift and nonlinear saturation generate a multibranch structure with two distinct transition routes.
In the linearly stable regime, a finite perturbation drives a first-order transition with type-I logarithmic scaling near threshold.
Beyond the linear onset, small perturbations grow tachyonically, and the response time controls both the growth and the subsequent breathing dynamics.
These results identify a common mechanism for scalarization across relativistic compact objects and nonrelativistic condensates, and point toward laboratory analogues in coherent media.
\end{abstract}

\maketitle

\textit{Introduction---}Scalarization is best known in compact objects in extensions of gravity, where matter or spacetime curvature drives an additional scalar field into a nonzero configuration~\cite{Damour:1993hw,Kleihaus:2015iea,Silva:2017uqg,Doneva:2017bvd}.
Recent studies have revealed a rich nonlinear transition structure in scalarization. Metastability and first-order transitions connect hairless and scalarized states~\cite{Unluturk:2025zie,Muniz:2025egq,Huang:2025dgc,Herdeiro:2026sur}, while dynamical evolution reveals spontaneous scalarization in boson stars~\cite{Alcubierre:2010ea} and critical behavior near threshold~\cite{Zhang:2021nnn}.
These developments raise a broader question. Is compact-object scalarization specific to relativistic strong gravity, or does it reflect a general mechanism by which coherent matter drives a coupled field to finite amplitude?

Self-gravitating nonrelativistic bosonic condensates provide a clean setting for this question.
Their Gross-Pitaevskii-Poisson (GPP) dynamics describe Newtonian boson stars and ultralight-dark-matter solitons~\cite{Liebling:2012fv,Schive:2014dra,Hui:2016ltb}.
In such systems, high density and macroscopic coherence can make weak local couplings dynamically important, as illustrated more broadly by medium-induced scalar responses in multicomponent dark sectors~\cite{Ferrante:2025avp}.
Collective responses can also propagate on medium-controlled timescales in superfluid dark matter and multicomponent Bose condensates~\cite{Berezhiani2015,RecatiStringari2022}, with two sound modes observed directly in a binary superfluid gas~\cite{KimHongShin2020}.
Together, these examples motivate a GPP condensate locally coupled to a scalar response field with an independent response timescale.

To isolate the mechanism within this setting, we study a self-gravitating GPP condensate locally coupled to a real scalar response field.
This model retains Newtonian self-gravity while allowing the effective-mass shift, nonlinear saturation, and response time to be controlled independently.
To our knowledge, it is the first explicit realization of scalarization in a self-gravitating GPP condensate.
The static branches and real-time evolution then provide a controlled test of how a common local mechanism organizes branch selection and critical behavior.

\textit{Model---}We consider a condensate wave function $\psi$, a Newtonian potential $\Phi$, and a real scalar response field $\varphi$ governed by the dimensionless equations
\begin{equation}
\begin{aligned}
i\partial_t\psi
&=-\frac12\nabla^2\psi+[\Phi+a(\varphi)]\psi,\\
(\nabla^2-\epsilon^2\partial_t^2)\varphi
&=4\pi\frac{da}{d\varphi}|\psi|^2,\\
\nabla^2\Phi
&=4\pi|\psi|^2.
\end{aligned}
\label{eq:model}
\end{equation}
The dimensional low-energy action and normalization leading to Eq.~\eqref{eq:model} are given in the End Matter.
With our normalization, the conserved scaled condensate mass is $M_b/M_0=\int d^3x\,|\psi|^2$, where $M_b$ is the total boson rest mass and $M_0$ is the corresponding mass scale.
The response-time parameter $\epsilon=v_0/c_\chi$ compares the characteristic condensate velocity $v_0$ with the propagation speed $c_\chi$ of the scalar response field.
The latter is set by the medium and need not equal the vacuum speed of light~\cite{Son:2002zn,Berezhiani2015}.
For $\epsilon>0$, the scalar response retains its own dynamics, whereas $\epsilon=0$ gives a quasi-static elliptic response.

We choose the even local coupling~\cite{Doneva:2017bvd}
\begin{equation}
a(\varphi)=\frac{\alpha_0}{2\beta}
\left(1-e^{-\beta\varphi^2}\right).
\label{eq:coupling}
\end{equation}
Unless stated otherwise, we use $\alpha_0=-10$ and $\beta=40$.
Near the bifurcation from the hairless branch, $a(\varphi)=(\alpha_0/2)\varphi^2-(\alpha_0\beta/4)\varphi^4+O(\varphi^6)$.
With $\alpha_0<0$ and $\beta>0$, the quadratic term controls the tachyonic onset, while the quartic term provides the leading nonlinear saturation that stabilizes finite-amplitude branches.
The same local mechanism applies to any even coupling with this low-field structure.
The coupling-function comparison, energetic definitions, and numerical implementation are also given in the End Matter.

\begin{figure*}[t]
\centering
\begin{overpic}[width=0.38\textwidth]{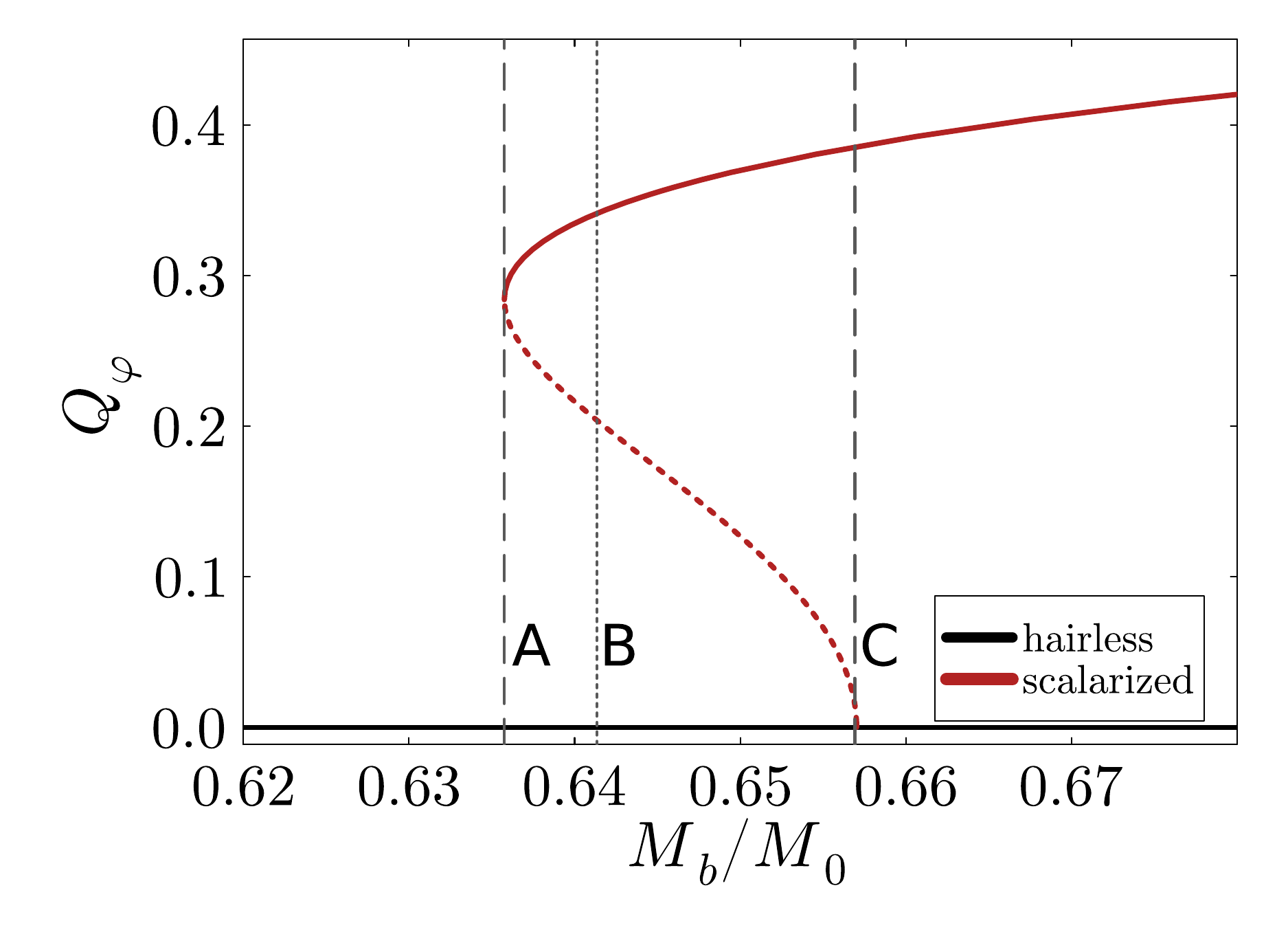}
\put(11,73){\normalsize (a)}
\end{overpic}\hspace{0.04\textwidth}
\begin{overpic}[width=0.38\textwidth]{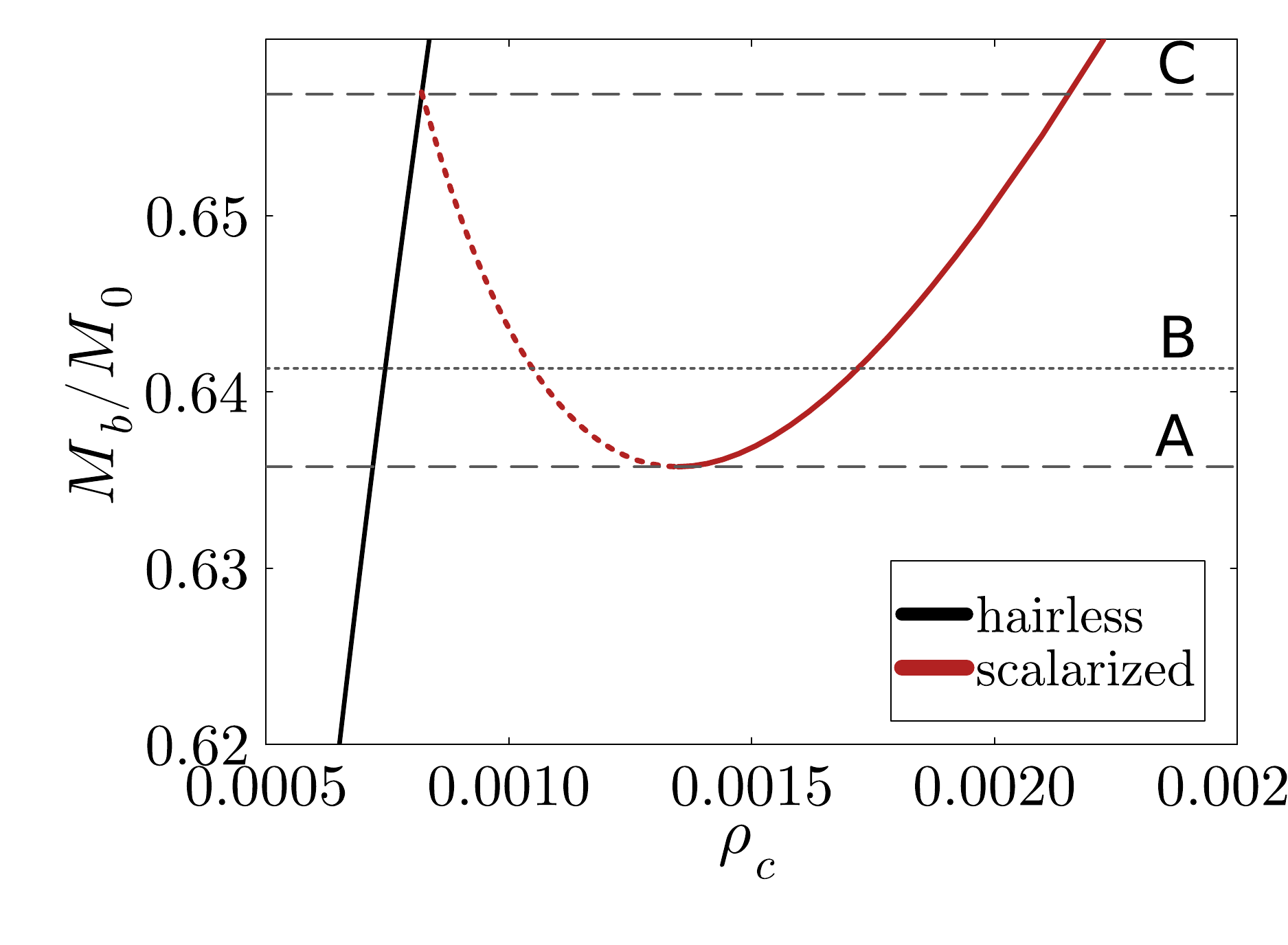}
\put(11,73){\normalsize (b)}
\end{overpic}\\[-0.5ex]
\begin{overpic}[width=0.38\textwidth]{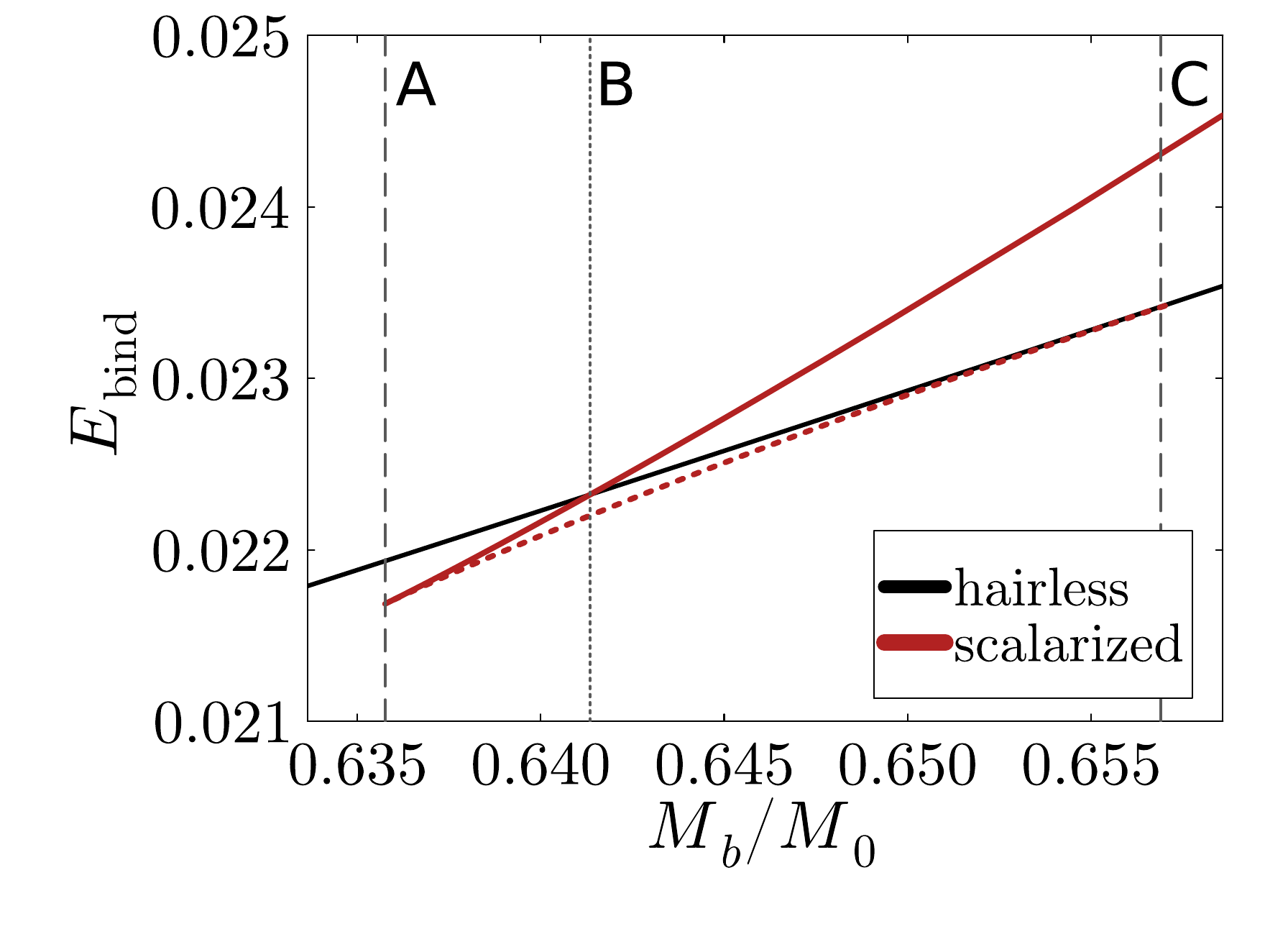}
\put(11,73){\normalsize (c)}
\end{overpic}\hspace{0.04\textwidth}
\begin{overpic}[width=0.38\textwidth]{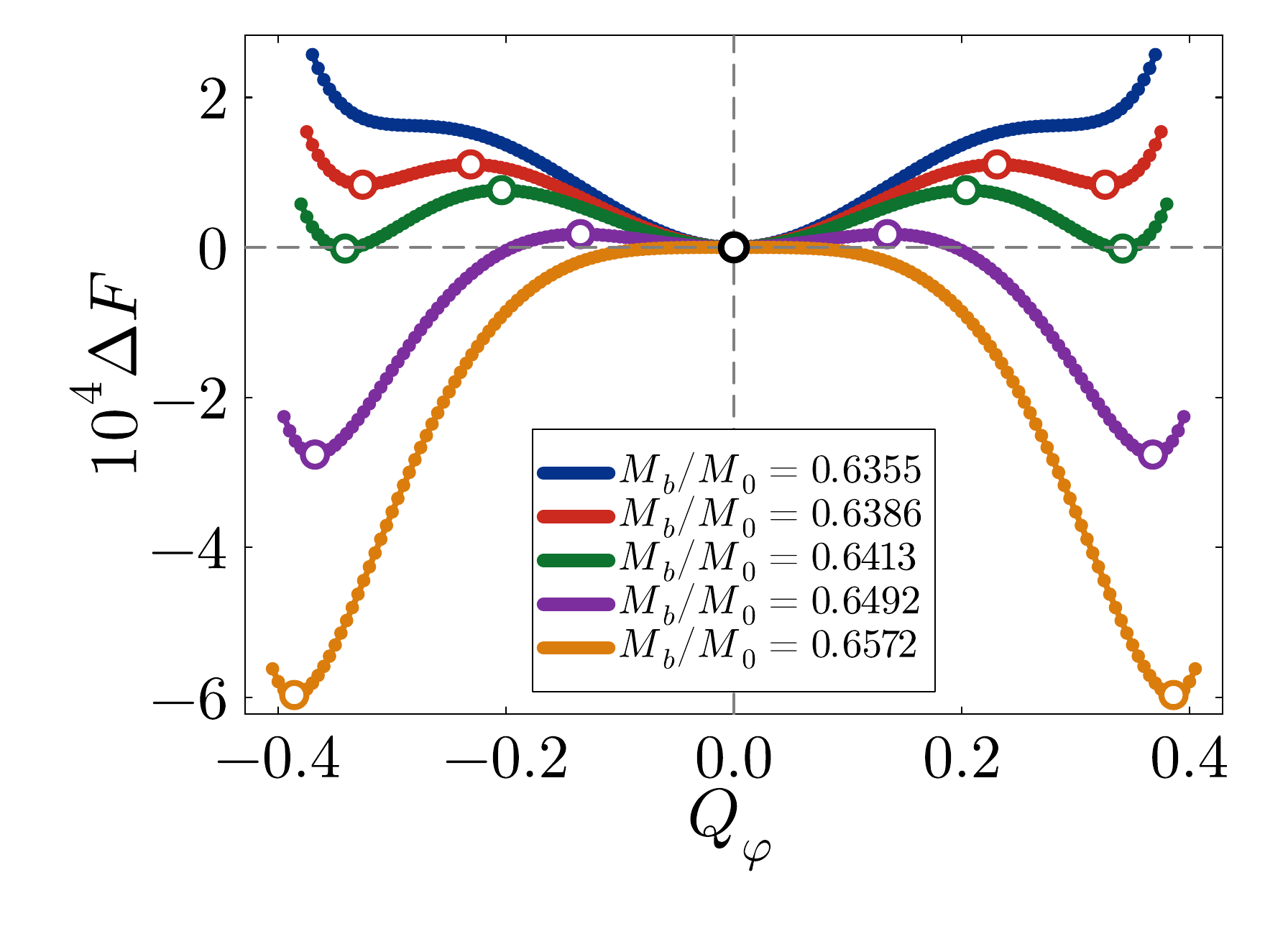}
\put(11,73){\normalsize (d)}
\end{overpic}
\caption{Static branches and fixed-mass energetics for $\alpha_0=-10$ and $\beta=40$.
(a) Scalar charge along the hairless and scalarized branches.
(b) Condensate mass as a function of central density.
Dotted segments have $dM_b/d\rho_c<0$.
(c) Binding energy of the hairless and scalarized branches.
(d) Fixed-mass constrained-energy difference $\Delta F$ for five representative masses.
Open circles mark the physical solutions with $\varphi_\infty=0$, and the negative-$Q_\varphi$ side follows from the $\varphi\to-\varphi$ symmetry.
The characteristic masses are $A=0.6358$, $B=0.6413$, and $C=0.6569$ in units of $M_b/M_0$.}
\label{fig:static}
\end{figure*}

\textit{Static branches and energetics---}Stationary solutions are regular and spherically symmetric, with $\psi(t,r)=e^{-i\omega t}\psi_0(r)$ and time-independent $\Phi$ and $\varphi$.
Figure~\ref{fig:static} summarizes the static solution structure.
Solutions with $\varphi=0$ form the hairless branch, which coincides with the ordinary GPP sequence.
Since $a(\varphi)$ is even, each nonzero solution has a symmetry-related partner under $\varphi\to-\varphi$, forming a pair of scalarized branches.
Over the multibranch interval, hairless and scalarized solutions coexist at the same $M_b/M_0$, allowing their energetics to be compared directly.

Panel (a) shows the scalar charge $Q_\varphi$ and the folded scalarized branch, a structure also found in relativistic compact objects~\cite{Kleihaus:2011sx,Silva:2018qhn,Herdeiro:2026sur}.
Panel (b) gives $M_b/M_0$ as a function of the condensate central density $\rho_c=|\psi_0(0)|^2$.
Segments with $dM_b/d\rho_c<0$ are shown dotted and are expected to be radially unstable according to the turning-point criterion~\cite{Kleihaus:2011sx,Huang:2025dgc}.
Panel (c) compares the energetics of the branches through the binding energy $E_{\rm bind}=-E_{\rm NR}/(M_b/M_0)$, where $E_{\rm NR}$ is the reduced nonrelativistic energy, as in related analyses of scalarized stars~\cite{Unluturk:2025zie,Muniz:2025egq,Huang:2025dgc}.
Taken together, panels (a)-(c) establish the sequence that organizes the dynamics.
The low- and high-density scalarized branches meet at the fold point A.
The high-density branch becomes globally favored at fixed mass after its binding energy crosses that of the hairless branch at B.
The hairless branch remains linearly stable up to C and becomes tachyonically unstable beyond it.
Between A and C, scalarized states therefore coexist with a linearly stable hairless state, so scalarization requires a finite perturbation.
To expose the barrier between competing branches, we allow an auxiliary asymptotic scalar value $\varphi_\infty$ and construct the fixed-mass constrained energy shown in panel (d).
For a massless scalar, $\varphi=\varphi_\infty+Q_\varphi/r+\cdots$ at large radius~\cite{Damour:1993hw,Muniz:2025egq}.
We define $F(Q_\varphi;M_b)=E_{\rm NR}+\varphi_\infty Q_\varphi$ and plot $\Delta F=F-F_0$, where $\varphi_\infty Q_\varphi$ is the boundary Legendre term and $F_0$ is the hairless-branch value at the same mass.
The physical solutions are recovered as stationary points at $\varphi_\infty=0$.
The resulting curves show the local minima and the nonlinear barrier separating the hairless and scalarized states.

\begin{figure*}[t]
\centering
\raisebox{-3pt}{\begin{overpic}[width=0.46\textwidth]{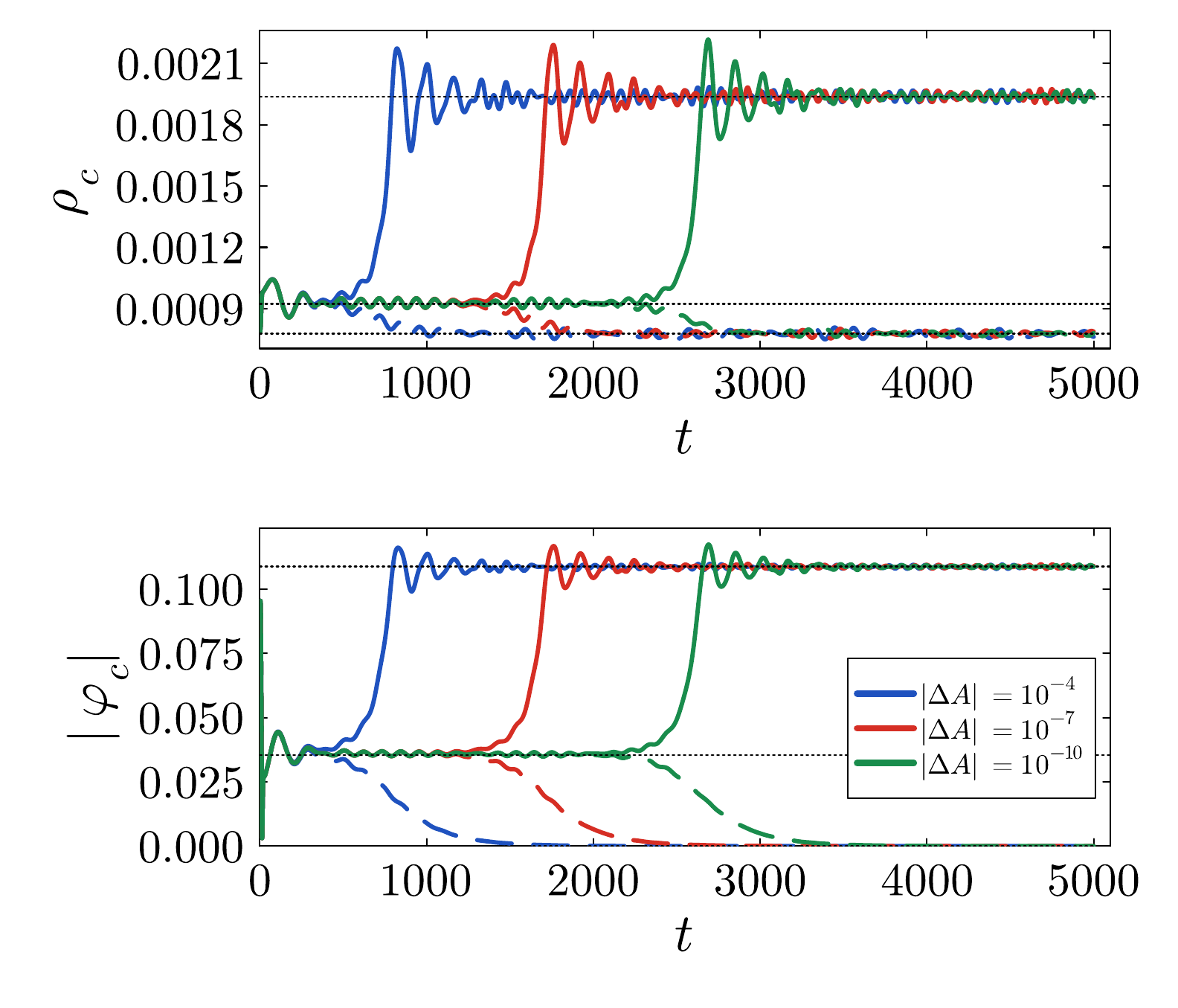}
\put(3,82){\normalsize (a)}
\end{overpic}}\hspace{0.005\textwidth}
\begin{overpic}[width=0.5\textwidth]{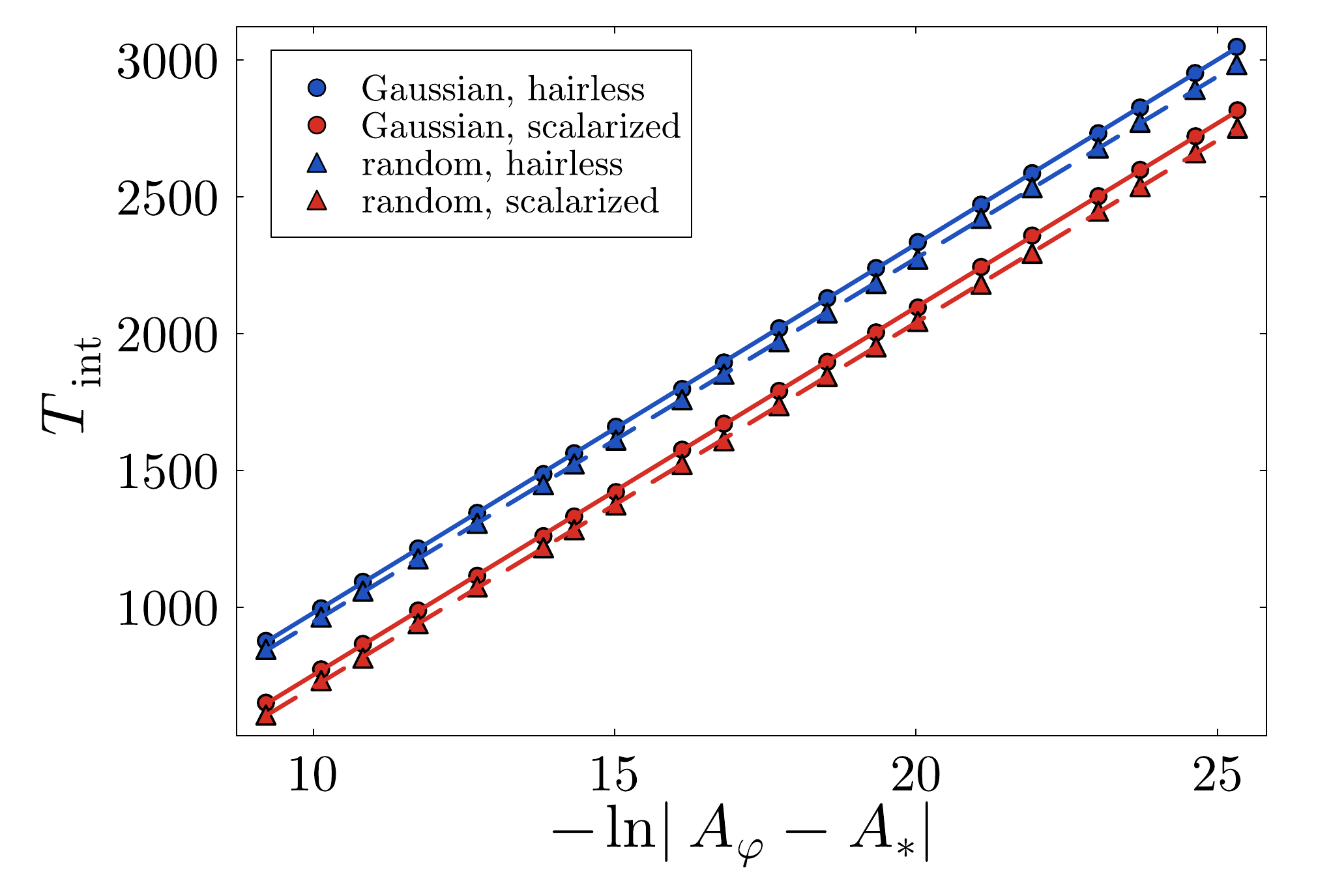}
\put(3,64){\normalsize (b)}
\end{overpic}
\caption{Finite-amplitude critical dynamics at $M_b/M_0=0.6492$ and $\epsilon=1$.
(a) Evolutions on opposite sides of the lower threshold approach the same long-lived intermediate state before departing toward hairless or scalarized final states.
Here $\Delta A=A_\varphi-A_*$.
Solid and dashed curves lead to scalarized and hairless final states, respectively.
Dotted lines indicate the central values of the matched static solutions associated with the hairless endpoint, intermediate state, and scalarized endpoint.
(b) For Gaussian and fixed smooth-random initial profiles, data on both sides of their respective thresholds $A_*$ follow the type-I lifetime law, with $\kappa^{-1}=134\pm1$.}
\label{fig:finite-amplitude}
\end{figure*}

The fixed-mass constrained energy is even in $Q_\varphi$ under the $\varphi\to-\varphi$ symmetry.
Its expansion about $Q_\varphi=0$ therefore takes the form
\begin{equation}
\Delta F(Q_\varphi;M_b)
=\frac{\lambda}{2}Q_\varphi^2
+\frac{u}{4}Q_\varphi^4
+\frac{v}{6}Q_\varphi^6+\cdots,
\label{eq:landau}
\end{equation}
where $\lambda$, $u$, and $v$ depend on $M_b$.
The quadratic coefficient $\lambda$ controls the linear stability of the hairless branch. It vanishes at C and becomes negative in the tachyonic regime.
The nonlinear coefficients $u$ and $v$ reflect both the low-field expansion of $a(\varphi)$ and the self-consistent readjustment of the condensate profile and Newtonian potential.
The structure in Fig.~\ref{fig:static}(d) is captured by $u<0$ and $v>0$.
The positive sixth-order term is the lowest-order contribution that bounds the expansion when the quartic coefficient is negative.
A finite-$Q_\varphi$ minimum and an intervening maximum can then appear while $\lambda>0$, producing the barrier for finite-amplitude scalarization.
When $\lambda$ changes sign, the hairless state becomes tachyonically unstable.
The two scalarization routes are therefore organized by the same low-order fixed-mass energy expansion.

\begin{figure}[!t]
\centering
\includegraphics[width=\columnwidth]{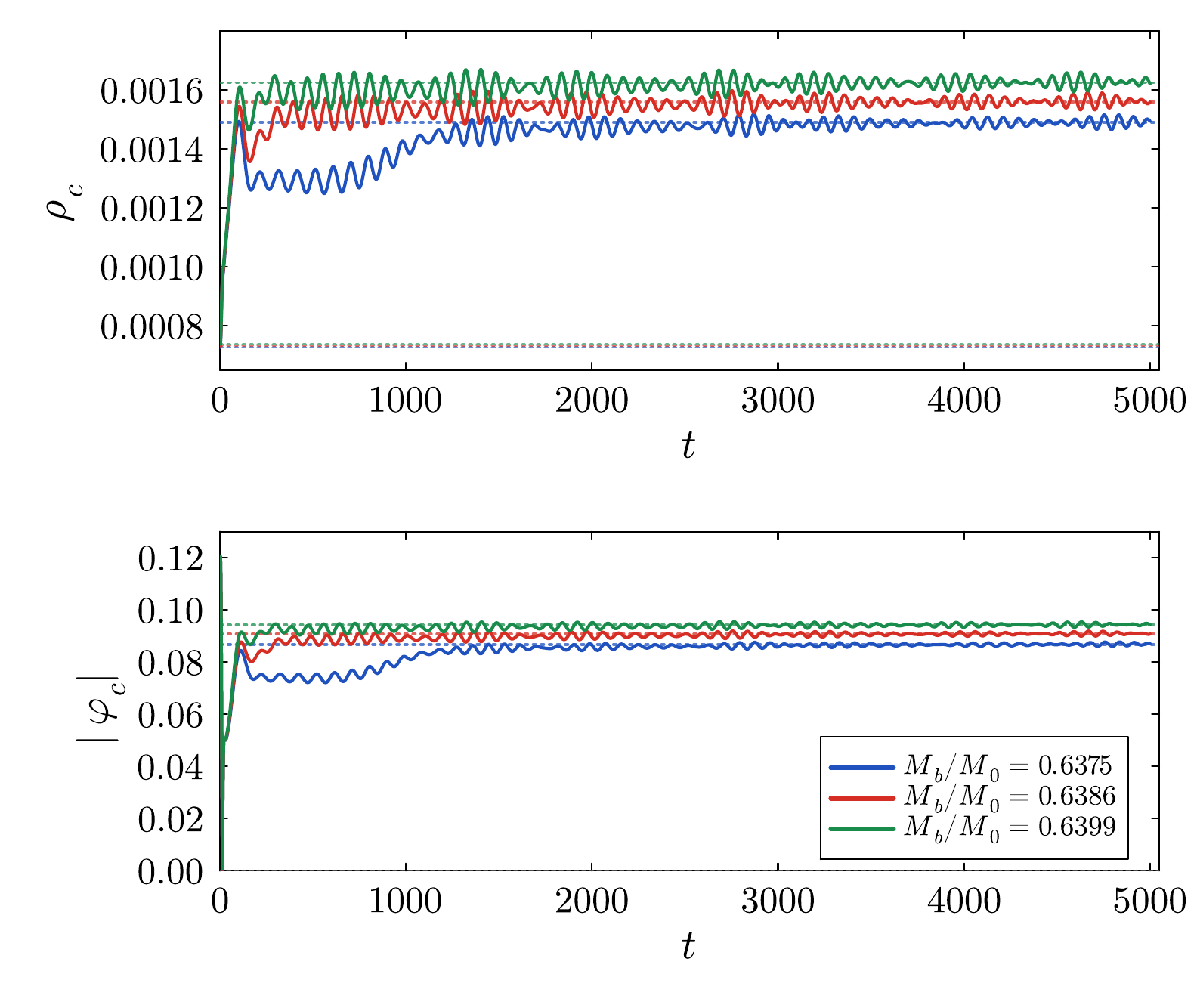}
\caption{Finite-amplitude scalarization from the hairless branch into long-lived metastable remnants at three masses between A and B.
Dotted lines mark the hairless background values and those of the static scalarized solutions matched to the late-time core profiles.
The matched core mass is slightly below the conserved total mass, with the remainder forming a diffuse outer component.
The envelope modulation arises from coherent beating between two long-lived mixed radial modes.}
\label{fig:metastable-transition}
\end{figure}

\textit{Dynamical transitions---}Finite-amplitude scalarization in the coexistence region is a first-order transition across the nonlinear barrier in Fig.~\ref{fig:static}(d).
Because the hairless branch remains linearly stable, the transition requires a finite perturbation.
Direct evolutions between A and B show that a sufficiently large perturbation can carry the condensate across the barrier from the globally favored hairless state into a long-lived metastable scalarized remnant (see Fig.~\ref{fig:metastable-transition}).
Across the finite-amplitude regime between A and C, the resulting scalarized states exhibit a common late-time structure.
They contain a scalarized core whose mass is slightly below the conserved total condensate mass and a diffuse outer component.
The nonadiabatic transition also excites two long-lived mixed radial modes, whose coherent interference produces beating in both $\rho_c$ and $|\varphi_c|$.

\begin{figure*}[t]
\centering
\begin{overpic}[width=0.46\textwidth]{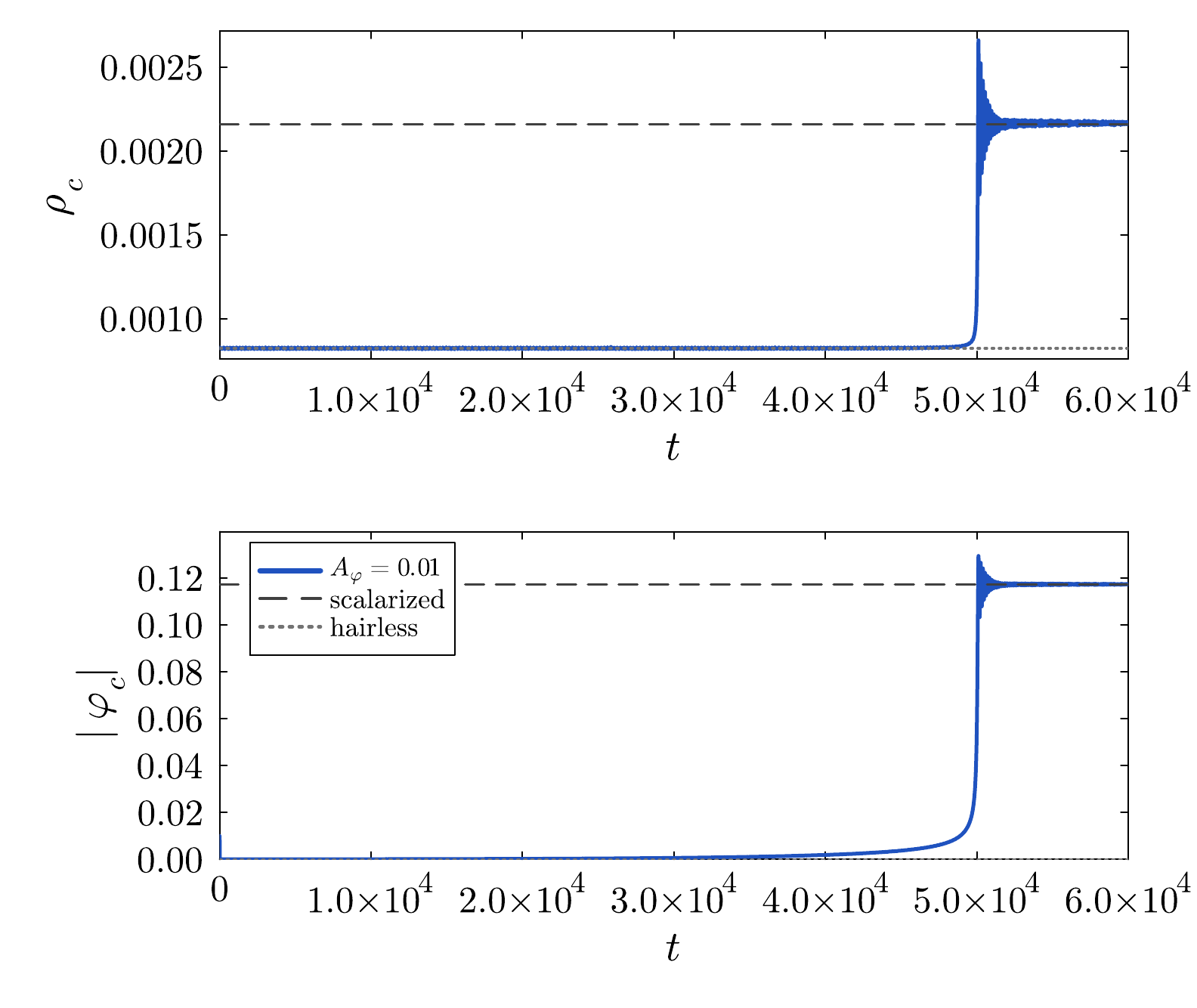}
\put(3,85){\normalsize (a)}
\end{overpic}\hspace{0.005\textwidth}
\begin{overpic}[width=0.47\textwidth]{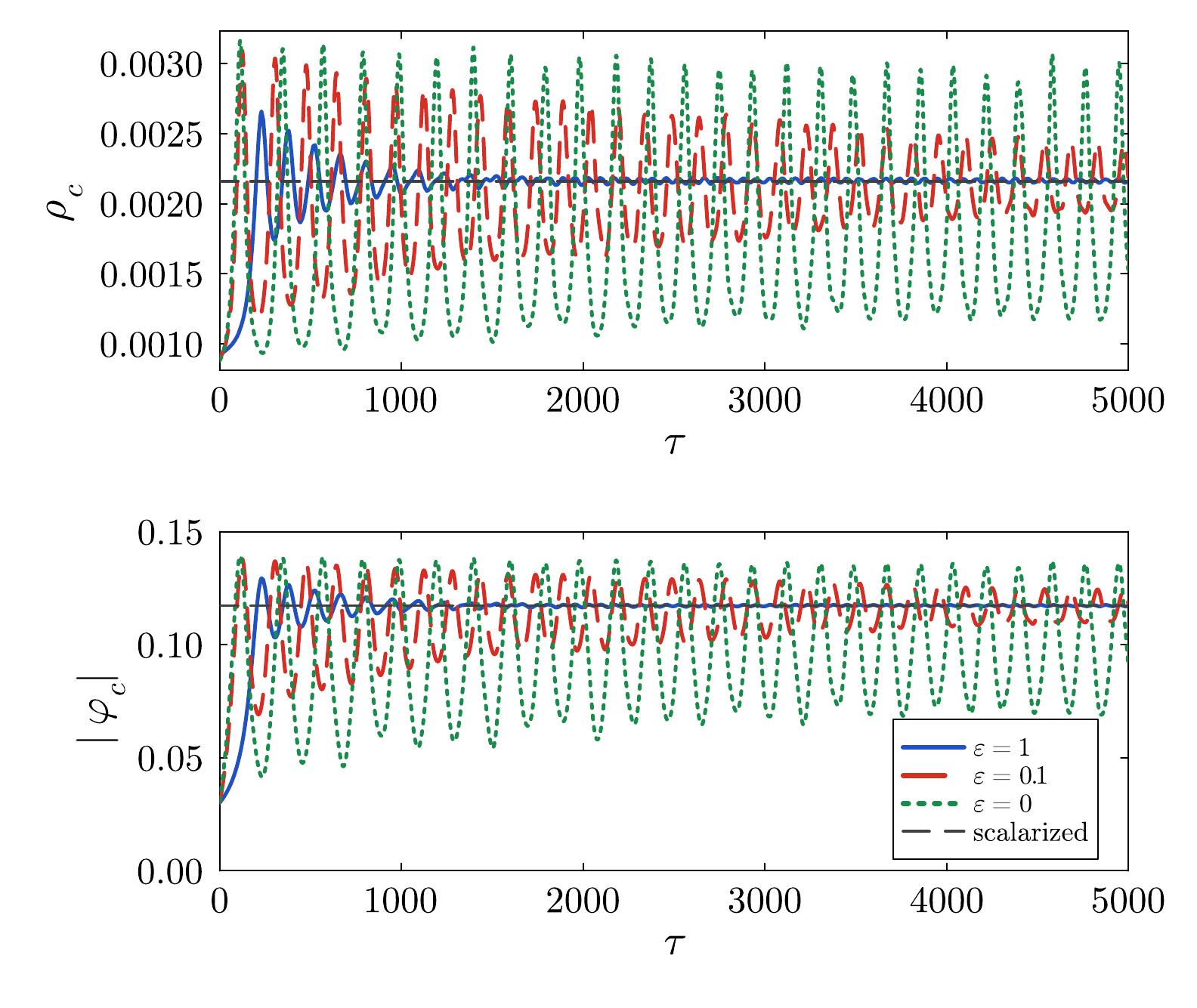}
\put(3,85){\normalsize (b)}
\end{overpic}
\caption{Scalarization dynamics above the linear onset at $M_b/M_0=0.6572$.
(a) Evolution from a Gaussian scalar perturbation with $A_\varphi=0.01$ toward a high-density scalarized breathing state for $\epsilon=1$.
Dashed and dotted lines mark the central values of the high-density scalarized and hairless static solutions, respectively.
(b) Late-time breathing for the same initial profile at $\epsilon=1$, $0.1$, and in the quasi-static limit $\epsilon=0$.
The time coordinate is $\tau=t-t_{\rm align}$, where $t_{\rm align}$ is the first upward crossing of $|\varphi_c|=0.03$.
Both the amplitude and period of the breathing depend on the response time.}
\label{fig:tachyonic}
\end{figure*}

For $M_b/M_0=0.6492$ and $\epsilon=1$, we take the initial scalar profile to be $\varphi(0,r)=A_\varphi\exp[-r^2/(2\sigma^2)]$, with $\sigma=8$ and $\partial_t\varphi(0,r)=0$.
Scalarized remnants form only over a finite range of $A_\varphi$, whose lower threshold $A_*$ separates relaxation to the hairless branch from evolution toward a scalarized remnant.
The upper boundary has a different origin.
As $A_\varphi$ increases, the coupling contribution to the initial energy approaches its finite bound, while the scalar-gradient contribution continues to grow.
Sufficiently strong perturbations therefore no longer evolve toward a scalarized state.
Near the lower threshold $A_*$, evolutions on either side approach the same long-lived intermediate state before departing toward the hairless or scalarized final state [Fig.~\ref{fig:finite-amplitude}(a)].
The lifetime of the intermediate state follows
\begin{equation}
T_{\rm int}\simeq-\frac{1}{\kappa}\ln|A_\varphi-A_*|+T_0,
\label{eq:lifetime}
\end{equation}
where $\kappa$ is the growth rate along its unstable direction.
This logarithmic divergence is characteristic of type-I critical behavior~\cite{Hawley:2000dt,Gundlach:2007gc}.
Similar logarithmic scaling appears in dynamical scalarization~\cite{Zhang:2021nnn}.
Although $A_*$ depends on the initial profile, fits on both sides of the threshold give $\kappa^{-1}\simeq134\pm1$ with $R^2>0.9999$ for both Gaussian and fixed smooth-random profiles [Fig.~\ref{fig:finite-amplitude}(b)].
The smooth-random profile is defined in the End Matter.
The intermediate state breathes around the low-density scalarized solution, and its logarithmic lifetime scaling is consistent with a single unstable mode, as expected for type-I criticality.
The finite-amplitude crossing found here relies on the independent dynamics of the response field.
Decreasing $\epsilon$ reduces the propagation timescale of the response field and narrows the transition window, rather than simply accelerating the same crossing process.
At $\epsilon=0$, $\varphi$ is determined on each time slice by an elliptic constraint sourced by the condensate, and the transition window closes for the initial profiles considered here.

Beyond the linear onset, the hairless branch has an unstable scalar mode.
For $\epsilon>0$, a localized scalar perturbation excites this mode and grows exponentially until nonlinear saturation drives the system toward a long-lived scalarized breathing state [Fig.~\ref{fig:tachyonic}(a)].
Linear theory gives $\gamma\propto\epsilon^{-1}$, a scaling reproduced by the evolutions.
Beyond the linear stage, backreaction couples the response field to the intrinsic condensate dynamics, so varying $\epsilon$ changes both the amplitude and period of the resulting breathing [Fig.~\ref{fig:tachyonic}(b)].
By contrast, at $\epsilon=0$, the scalar response is determined elliptically on each time slice, corresponding to quasi-static branch selection.

\textit{Discussion---}Our results establish a condensate-induced effective-mass shift and nonlinear saturation as a minimal mechanism for scalarization in a nonrelativistic self-gravitating condensate.
Together, these ingredients produce a regime in which hairless and scalarized branches coexist.
The resulting fixed-mass energy structure supports a first-order, finite-amplitude transition while the hairless state remains linearly stable, and tachyonic scalarization after that state becomes linearly unstable.
The lifetime near the finite-amplitude threshold obeys the type-I logarithmic law.
The transition leaves a scalarized core whose mass is slightly below the conserved total mass, together with a diffuse outer component.
It also excites two long-lived mixed radial modes whose interference produces coherent beating.
If this beating survives in a relativistic extension, it could appear as a slow envelope in periodic gravitational-lensing signals from oscillating boson stars~\cite{Yang:2025yej}.
Across both routes, the response time controls the finite-amplitude window, tachyonic growth, and subsequent breathing dynamics.

Local coupling, nonlinear saturation, and finite response are common to relativistic compact objects, nonrelativistic dark-sector condensates, and multicomponent coherent media.
Our results therefore place scalarization within a broader class of coupled-field transitions and motivate laboratory analogues in coherent-wave systems.

\textit{Acknowledgments---}We gratefully acknowledge discussions with J. Kunz.
This work is supported by the National Natural Science Foundation of China (Nos.12305064, 12665010, 12365009, 12405064 and 12565010) and the Jiangxi Provincial
Natural Science Foundation (Nos.20242BCE50055, 20262BAC240347, 20262BAC240348).
Y.~S.~Myung is supported by the National Research Foundation of Korea (NRF) grant funded by the Korea government (MSIT) (RS-2022-NR069013).

\bibliography{references}

\onecolumngrid
\begin{center}
    \vspace{1em}
    {\large\bfseries End Matter \par}
    \vspace{1em}
\end{center}
\twocolumngrid

\begin{figure*}[t]
\centering
\includegraphics[width=\textwidth]{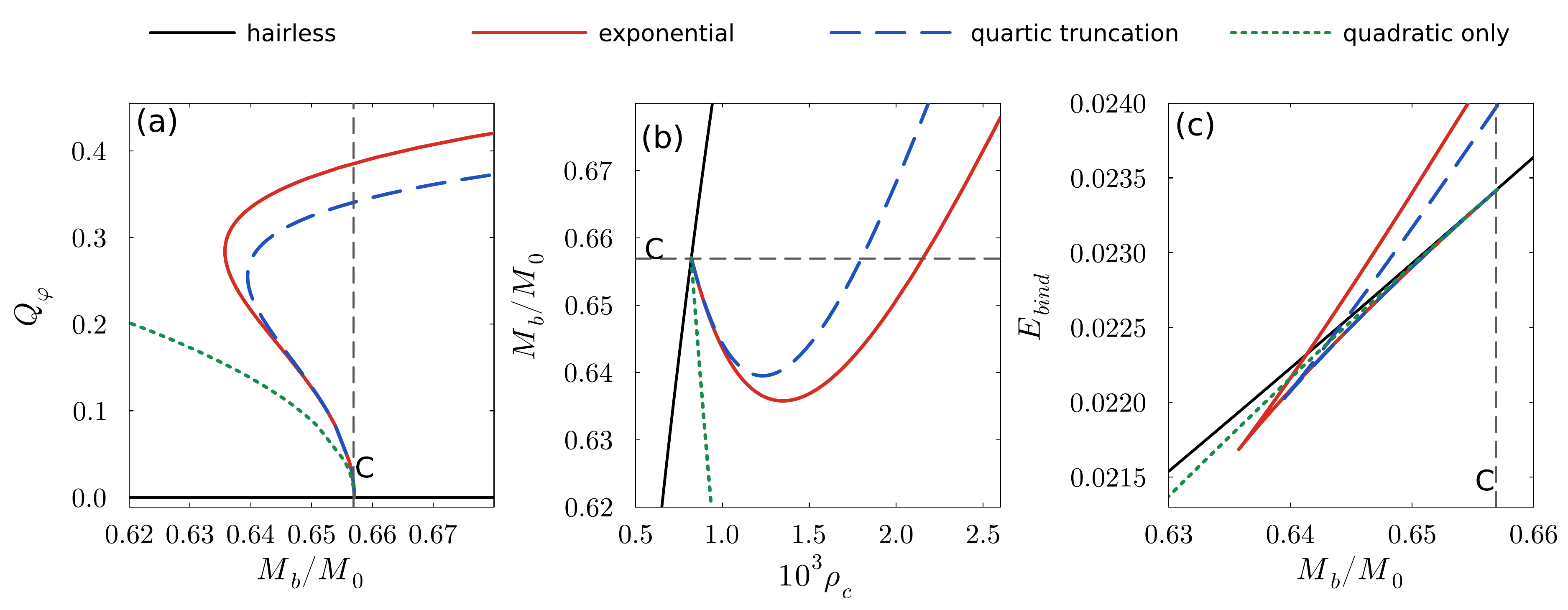}
\caption{Role of nonlinear saturation in the static branch structure for three coupling functions with the same quadratic coefficient.
Black denotes the hairless branch.
Red, blue, and green denote the exponential coupling, its quartic truncation, and the quadratic-only coupling, respectively.
(a) Scalar charge as a function of condensate mass.
All three scalarized sequences share the same linear onset, while only the couplings with nonlinear saturation develop a folded scalarized branch.
(b) Condensate mass as a function of central density.
(c) Binding-energy ordering near the bifurcation region.
The exponential and quartic-truncated models exhibit an energy crossing with the hairless branch below the common onset, whereas the quadratic-only model shows no separate crossing or exchange of energetic ordering.}
\label{fig:coupling-robustness}
\end{figure*}

\textit{Dimensional action and normalization---}In units with $\hbar=c=1$, a dimensional low-energy action that reduces to Eq.~\eqref{eq:model} is
\begin{equation}
\begin{aligned}
S=\int dt\,d^3x\,\Bigg[&
\frac{i}{2}(\psi^*\dot\psi-\psi\dot\psi^*)
-\frac{|\nabla\psi|^2}{2m}
-m\Phi|\psi|^2\\
&+\frac{\dot\chi^2}{2c_\chi^2}
-\frac{|\nabla\chi|^2}{2}
-\frac{\Delta M^2(\chi)}{2m}|\psi|^2\\
&-\frac{|\nabla\Phi|^2}{8\pi G}\Bigg].
\end{aligned}
\label{eq:dimensional-action}
\end{equation}
Here $m$ is the condensate particle mass, $|\psi|^2$ is its number density, and $\chi$ is a real response field with propagation speed $c_\chi$ in the medium.
We take $\Delta M^2(\chi)=-\mu^2[1-\exp(-\chi^2/f_\chi^2)]$.
At low field, $\Delta M^2=g_2\chi^2+c_4\chi^4+O(\chi^6)$, where $g_2=-\mu^2/f_\chi^2<0$ provides the density-dependent negative contribution that can destabilize the hairless state, while $c_4=\mu^2/(2f_\chi^4)>0$ gives the leading nonlinear saturation.

Let $L_0=(mv_0)^{-1}$ and $T_0=(mv_0^2)^{-1}$, and write $\bm{x}=L_0\hat{\bm{x}}$, $t=T_0\hat t$, $\Phi=v_0^2\hat\Phi$, $\psi=v_0^2\sqrt{m/G}\,\hat\psi$, and $\chi=v_0^2\hat\varphi/\sqrt{4\pi G}$.
With $a(\hat\varphi)=\Delta M^2(v_0^2\hat\varphi/\sqrt{4\pi G})/(2m^2v_0^2)$, variation of Eq.~\eqref{eq:dimensional-action} with respect to $\psi^*$, $\chi$, and $\Phi$, followed by these rescalings, gives Eq.~\eqref{eq:model} after the hats are dropped.
Matching Eq.~\eqref{eq:coupling} gives
\begin{equation}
\begin{aligned}
\epsilon&=\frac{v_0}{c_\chi},\\
\mu^2&=-\frac{\alpha_0}{\beta}m^2v_0^2,
\qquad
f_\chi=\frac{v_0^2}{\sqrt{4\pi G\beta}},\\
g_2&=\frac{4\pi Gm^2}{v_0^2}\alpha_0,
\qquad
c_4=-\frac{8\pi^2G^2m^2}{v_0^6}\alpha_0\beta.
\end{aligned}
\label{eq:parameter-mapping}
\end{equation}
For $\alpha_0=-10$, $\beta=40$, $m=10^{-22}\,\mathrm{eV}$, and $v_0/c=3.34\times10^{-4}$, the benchmark parameters are $\mu\simeq1.67\times10^{-26}\,\mathrm{eV}$, $f_\chi\simeq6.06\times10^{10}\,\mathrm{GeV}$, and $g_2\simeq-7.58\times10^{-92}$.
The maximal mass-squared shift is $\mu^2/m^2\simeq2.8\times10^{-8}$, while its one-particle energy scale is $\mu^2/(2m)=0.125\,mv_0^2$.
For representative cores, $|g_2||\psi|^2L_0^2/m=4\pi|\alpha_0||\hat\psi|^2$ lies between $0.10$ and $0.27$.
The interaction is therefore weak relative to the rest-mass scale but dynamically significant on the condensate scale.

\textit{Fixed-mass energetics---}For a static isolated configuration, the nonrelativistic energy and binding energy per particle are
\begin{equation}
\begin{aligned}
E_{\rm NR}={}&\int_{r\leq R}d^3x\bigg[
\frac12|\nabla\psi|^2+\frac12\Phi|\psi|^2\\
&\quad
+a(\varphi)|\psi|^2
+\frac{|\nabla\varphi|^2}{8\pi}\bigg]
+\frac{Q_\varphi^2}{2R},\\
E_{\rm bind}={}&-\frac{E_{\rm NR}}{N},
\qquad N=\frac{M_b}{M_0}.
\end{aligned}
\label{eq:energetics}
\end{equation}
The dimensional boson rest mass is $M_b=mN_{\rm phys}$, where $N_{\rm phys}$ is the particle number.
In a relativistic boson-star description, $E_{\rm NR}$ is the leading nonrelativistic part of $M_{\rm ADM}-M_b$, while $Q_\varphi$ plays the role of the asymptotic scalar charge~\cite{Whinnett:1999sc,Liebling:2012fv,Huang:2025dgc}.
The last term accounts for the scalar-gradient energy outside the numerical domain, where the massless response has the asymptotic form $\varphi(r)=\varphi_\infty+Q_\varphi/r+\cdots$.

To construct Fig.~\ref{fig:static}(d), we allow $\varphi_\infty$ to act as an auxiliary source while holding $N$ fixed.
Along the static solution family, the scalar boundary variation is $\delta E_{\rm NR}=-Q_\varphi\,\delta\varphi_\infty$.
The corresponding fixed-mass constrained energy is therefore $F(Q_\varphi;M_b)=E_{\rm NR}+\varphi_\infty Q_\varphi$, for which $\delta F=\varphi_\infty\,\delta Q_\varphi$.
Figure~\ref{fig:static}(d) shows $\Delta F=F-F_0$, with $F_0$ evaluated on the hairless branch at the same mass.
Physical solutions have $\varphi_\infty=0$ and are stationary points of $F$.
At these points, defining $\Delta E_{\rm bind}=E_{\rm bind}-E_{{\rm bind},0}$, one has $\Delta E_{\rm bind}=-\Delta F/N$.

To isolate the roles of the quadratic coupling and nonlinear saturation, we compare the exponential coupling with its quadratic-only form $a_2=\alpha_0\varphi^2/2$ and quartic truncation $a_4=\alpha_0\varphi^2/2-\alpha_0\beta\varphi^4/4$ in Fig.~\ref{fig:coupling-robustness}.
All three couplings share the same quadratic coefficient and therefore the same linear onset.
The exponential and quartic-truncated couplings both develop a folded scalarized branch and an exchange of energetic ordering, whereas the quadratic-only coupling does not.
Thus, the common quadratic term fixes the tachyonic onset, while bounded nonlinear terms generate the folded multibranch structure that underlies finite-amplitude scalarization.

\textit{Numerical methods and checks---}All calculations assume spherical symmetry on a uniform radial grid.
Static branches are obtained by finite-difference discretization of the radial boundary-value equations followed by Newton iteration.
Regularity is imposed at the origin.
The outer conditions are $\psi(R)=0$, $\Phi(R)=-N/R$, and $\varphi'(R)+\varphi(R)/R=0$.
Time evolution uses Strang splitting for the condensate, a Crank--Nicolson radial kinetic step, and a Poisson solve on each time slice.
For $\epsilon>0$, the response field is advanced by a second-order leapfrog scheme with $\Delta t\lesssim\epsilon\Delta r$ and the outgoing condition $\epsilon\partial_t\varphi+\partial_r\varphi+\varphi/R=0$.
At $\epsilon=0$, it is solved elliptically.

The static branches are computed with $R=150$ and $\Delta r=0.1$.
Their residuals remain below $2\times10^{-7}$, with fixed-mass mismatches below $3\times10^{-6}$.
An independent scalar eigenvalue calculation gives $M_b/M_0\simeq0.6569$ for the linear onset, consistent with the branch bifurcation.
Critical evolutions use $R=3000$, $\Delta r=0.1$, and $\Delta t=0.01$, with all lifetime-fitting intervals ending before any boundary-reflected signal reaches the core.
Particle number is conserved to better than $10^{-9}$.
We use Gaussian and fixed smooth-random initial profiles.
The latter is
\begin{equation}
f_{\rm R}(r)=\mathcal N_{\rm R}e^{-r^2/(2\sigma_{\rm R}^2)}
\left[1+\eta\sum_{n=1}^{6}c_n
\cos\left(\frac{n\pi r}{R_{\rm c}}\right)\right],
\end{equation}
where $\sigma_{\rm R}=12$, $R_{\rm c}=48$, $\eta=0.42$,
$(c_1,c_2,c_3)=(0.154312,-0.250919,0.734566)$, and $(c_4,c_5,c_6)=(0.260497,-0.547442,-0.078028)$.
The profile is normalized to $\max|f_{\rm R}|=1$ and used as $\varphi(0,r)=A_\varphi f_{\rm R}(r)$ with $\partial_t\varphi(0,r)=0$.
For Gaussian and smooth-random profiles, fits over $10^{-11}\lesssim|A_\varphi-A_*|\lesssim10^{-4}$ give $\kappa^{-1}\simeq134\pm1$ with $R^2>0.9999$.
Across the tested box sizes, the relative changes in the growth rate and in each breathing frequency remain below $5\times10^{-7}$ and $10^{-3}$, respectively.

\end{document}